\documentclass[apj]{emulateapj}
\usepackage{natbib}
\usepackage{amsmath}
\usepackage{apjfonts}

\newcommand{\cm}{{\mathrm{cm}}}

\newcommand{\erg}{{\mathrm{erg}}}

\begin{document}

\title{Off-Axis Gamma-Ray Burst Afterglow Modeling Based On
  A Two-Dimensional Axisymmetric Hydrodynamics Simulation}

\author{Hendrik van Eerten, Weiqun Zhang and Andrew MacFadyen}
\affil{
  Center for Cosmology and Particle Physics, Physics
  Department, New York University, New York, NY 10003}

\begin{abstract}

  Starting as highly relativistic collimated jets, gamma-ray burst
  outflows gradually decellerate and become non-relativistic spherical
  blast waves.  Although detailed analytical solutions describing the
  afterglow emission received by an on-axis observer during both the
  early and late phases of the outflow evolution exist, a calculation
  of the received flux during the intermediate phase and for an
  off-axis observer requires either a more simplified analytical model
  or direct numerical simulations of the outflow dynamics. In this
  paper we present light curves for off-axis observers covering the
  long-term evolution of the blast wave, calculated from a high
  resolution two-dimensional relativistic hydrodynamics simulation
  using a synchrotron radiation model.  We compare our results to
  earlier analytical work and calculate the consequence of the
  observer angle with respect to the jet axis both for the detection
  of orphan afterglows and for jet break fits to the observational
  data. We find that observable jet breaks can be delayed for up to
  several weeks for off-axis observers, potentially leading to
  overestimation of the beaming corrected total energy. When using our
  off-axis light curves to create synthetic Swift X-ray data, we find
  that jet breaks are likely to remain hidden in the data. We also
  confirm earlier results in the literature finding that only a very
  small number of local Type Ibc supernovae can harbor an orphan
  afterglow.

\end{abstract}


\section{Introduction}
\label{sec:intro}

According to the standard fireball shock model, gamma-ray burst (GRB)
afterglows are the result of the interaction between a decelerating
relativistic jet and the surrounding medium. Synchrotron radiation is
produced by shock-accelerated electrons interacting with a
shock-generated magnetic field.  The radiation will peak at
progressively longer wavelengths and the observed light curve will
change shape whenever the observed frequency crosses into different
spectral regimes, when the flow becomes non-relativistic and when
lateral spreading of the initially strongly collimated outflow becomes
significant (see e.g. \citealt{Zhang2004, Piran2005, Meszaros2006} for
recent reviews).

Analytical models have greatly enhanced our understanding of GRB
afterglows.  Such models rely on a number of simplifications of the
fluid properties and radiation mechanisms involved. Both at early
relativistic and late non-relativistic stages spherical symmetry can
be assumed. At first lateral spreading of the jet has not yet set in
and the beaming is so strong that a collimated outflow is still
observationally indistinguishable from a spherical flow. Eventually
the outflow really has become approximately spherical.  Self-similar
solutions for a strong explosion can be applied, the Blandford-McKee
(BM, \citealt{Blandford1976}) solution in the relativistic regime, and
the Sedov-Taylor-von Neumann (ST, \citealt{Sedov1959, Taylor1950})
solution in the non-relativistic regime. However, in order to include
lateral spreading of the jet and to calculate the light curve for an
observer not located on the axis of the jet, the downstream fluid
profile has usually been approximated by a homogeneous slab (e.g.
\citealt{Rhoads1999, Kumar2000, Granot_etal_2002_ApJ, Waxman2004,
  Oren2004}). Structured jet models exist where the effect of
lateral expansion is estimated and fluid quantities like Lorentz
factor and density depend on the angle of the flow with respect to the
jet axis (e.g. \citealt{Rossi_PD_2008_MNRAS}, also
\citealt{Granot2007} and references therein).

However, to gain an understanding of afterglow light curves during all
stages of jet evolution and for off-axis observers, large scale
multi-dimensional simulations are required.  Over the past ten years
various groups have combined one-dimensional relativistic
hydrodynamics (RHD) simulations with a radiation calculation
(e.g. \citealt{Kobayashi1999, Downes2002, Mimica2009,
  vanEerten2010}). Thanks to specialized techniques such as
adaptive-mesh-refinement (AMR), jet simulations in more than one
dimension have also become feasible (e.g. \citealt{Granot2001,
  Zhang_MacFadyen_2006_ApJS, Meliani2007}).

In this work we present the results of a high-resolution
two-dimensional RHD simulation covering the full transition from
relativistic to non-relativistic flow. We use the simulation results
from \citealt{Zhang2009} (hereafter ZM09), but we now calculate for
the first time detailed light curves for off-axis observers. The
simulations have been performed with the \textsc{ram} code
\citep{Zhang_MacFadyen_2006_ApJS}.

Off-axis observations of GRB afterglows are observationally relevant
for a variety of reasons. Following the first observations of GRB
afterglows, it was immediately realized that, if the afterglow emission
was less strongly beamed than the prompt emission, many \emph{orphan}
afterglows should in principle still be observable even when the
prompt emission was not visible because the observer was positioned
too far away from the jet axis \citep{Rhoads1997, Rhoads1999}.
Detailed light curves from simulations help constrain the expected
rate of occurrence of orphan afterglows and help determine to what
extent orphan afterglows can possibly remain hidden in observations
of type Ibc supernovae.  This in turn helps to constrain the fraction of
type Ibc supernovae that can in principle be linked to GRBs.

For observers that are not too far off-axis but are still located
within the jet opening angle, the prompt emission still remains
visible and they will observe an afterglow light curve that shows a
jet break when the jet edges become visible, an effect which is
enhanced when lateral spreading becomes significant. The shape of this
jet break is expected to depend strongly on the observer angle. If the
observer is not on-axis, each edge of the jet becomes visible at
different times and the corresponding break is split in two, or
at least becomes smoother. This effect may account for the difficulty in
detecting a jet break for many GRBs \citep{Racusin2009, Evans2009}.

This paper is structured as follows. In section \ref{hydro_section} we
briefly review the RHD simulation setup and methods we have applied in
ZM09 and that form the basis for this paper as well. In section
\ref{radiation_section} we describe how the radiation is calculated
for an observer at an arbitrary angle with respect to the jet axis.
The resulting light curves are presented in section
\ref{results_section}. They are put into context first by a comparison
to light curves calculated from the relativistic BM solution for
on-axis observers, followed by a comparison to light curves from a
simplified homogeneous slab model for off-axis observers. We then
apply the simulation to two different observational issues. In section
\ref{jetbreaks_section} we use our computed light curves to generate
synthetic Swift data, to which we then fit broken and single power
laws in order to probe the extent to which X-ray jet breaks can be
hidden in the SWIFT data due to off-axis observer angle. In section
\ref{orphan_afterglows_section} we confirm the result from
\citet{Soderberg2006} that only a very small number of local Ibc
supernovae can possibly harbor an orphan GRB afterglow, now using our
simulation instead of a simplified analytical model for comparison
with the observations. We discuss the results presented in this paper
in section \ref{discussion_section}. The mathematical details of the
analytical model with which we have compared our simulation results
are summarized in the Appendix.

\section{Methods}
\subsection{Hydrodynamic Model}
\label{hydro_section}

We have used the two-dimensional RHD simulation first presented in
ZM09 as the basis for our calculations. This simulation was performed
using the \textsc{ram} adaptive mesh refinement code
\citep{Zhang_MacFadyen_2006_ApJS}.  R\textsc{am} employs the
fifth-order weighted essentially non-oscillatory (WENO) scheme
\citep{Jiang_Shu_1996_JCP} and uses the PARAMESH AMR tools
\citep{Macneice_etal_2000} from FLASH 2.3
\citep{Fryxell_etal_2000_ApJS}.

The simulation takes a conic section of the Blandford-McKee (BM)
analytical solution \citep{Blandford1976} as the initial condition,
starting from a fluid Lorentz factor directly behind the shock front
equal to 20. The isotropic energy of the explosion was set at $E_{iso}
= 10^{53}$ erg. The jet half opening angle $\theta_0 = 0.2$ rad
($11.5^\circ$), leading to a total energy in the twin jets of $E_j
\approx 2.0 \times 10^{51}$ erg. The circumburst proton number density
is taken to be homogeneous and set at $n = 1$ cm$^{-3}$. The pressure
$p$ of the surrounding medium is set at a very low value compared to
the density $\rho$ ($p = 10^{-10} \rho c^2$, with $c$ the speed of
light) and will therefore not be dynamically important. Under these
conditions, the starting radius of the blast wave is equal to $R_0
\approx 3.8 \times 10^{17}$ cm.

A spherical grid $(r, \theta)$ was used with $0 \le r \le 1.1 \times
10^{19}$ cm and $0 \le \theta \le \pi / 2$. At first, 16 levels of
refinement are used, with the finest cell having a size of $\Delta r
\approx 5.6 \times 10^{13}$ cm and $\Delta \theta \approx 9.6 \times
10^{-5}$ rad. The maximum refinement level is gradually decreased (but
always kept at least at 11) during the simulation, making use of the
fact that the blast wave widens proportionally to $t^4$ in lab frame
time to keep the number of cells radially resolving the blast wave
approximately constant.

The most important dynamical results from ZM09 are the following. They
find that very little sideways expansion takes place for the
ultrarelativistic material near the forward shock, while the mildly
relativistic and Newtonian jet material further downstream undergoes
more sideways expansion. When taking a fixed fraction of the total
energy contained within an opening angle as a measure of the jet
collimation it is found that sideways expansion is logarithmic (and
not exponential, as used by some early analytic models such as that of
\citealt{Rhoads1999}). This sideways expansion sets in approximately
at a time $t_\theta$ calculated from plugging $\gamma = 1 / \theta_j$
into the BM solution, where $\gamma$ the fluid Lorentz factor directly
behind the forward shock and $\theta_j$ the original jet opening
angle. For the simulation settings described above, $t_\theta \approx
373$ days, measured in the frame of the burster. The jet becomes
nonrelativistic and the BM solution breaks down at $t \backsim t_{NR}
\approx 970$ days. The time $t_{NR}$ is estimated by equating the
isotropic equivalent energy in the jet to the rest mass energy of the
material swept up by a spherical explosion, assuming the jet moves at
approximately the speed of light. The transition to spherical flow was
found to be a slow process and was found from the simulation to take
until $5 t_{NR}$ to complete.  After that time the outflow can be
described by the Newtonian Sedov-von Neumann-Taylor (ST) solution.

\subsection{Calculation of Off-Axis Afterglow Emission}
\label{radiation_section}
A large number of data dumps (2800) from the hydrodynamic simulation
are stored. These data dumps are then used to calculate the
synchrotron emission from all the individual fluid elements at the lab
frame time of each data dump. A single emission time corresponds to
different arrival times at the observer for different parts of the
fluid due to light travel time differences. The radiation from the
data dumps is binned over a number of observer times.  The size of
each fluid element is given by $\Delta V = r^2 \sin{\theta} \Delta r
\Delta \theta \Delta \phi$ in 3D. In the 2D axisymmetric simulation
the flow is independent of $\phi$ and, if the observer is positioned
on-axis, the $\phi$ symmetry allows for considering just the 2D
elements $\Delta V = 2 \pi r^2 \sin{\theta} \Delta r \Delta \theta$
provided by the data dumps. When calculating emission for off-axis
observers, however, fully 3D data must be created from the 2D fluid
elements by extending the data in the $\phi$ direction. The fluid
elements are split into smaller elements along the $\phi$ direction of
angular size $\Delta \phi$ to account for differences in relativistic
beaming and Doppler shifts of emission observed from different
angles. In practice we have started with angular resolution such that
$r \sin{\theta} \Delta \phi$ is comparable to $\Delta r$ and $r \Delta
\theta$, taking into account that the arrival time between the close
and far edge of the fluid element in the $\phi$ direction should stay
very small. We found that the light curves are not very sensitive to
the resolution in the $\phi$ direction, even when it is decreased
tenfold.

The synchrotron emission itself is calculated following
\citet{Sari1998}. We sum over the contributions of the individual
fluid elements. In the frame comoving with the fluid element, the
spectral power peaks at
\begin{equation}
 P' = 0.88 \left( \frac{16}{3} \right)^2 \frac{p-1}{3p-1} \frac{\sigma_T m_e
c^2}{8 \pi q_e} n' B',
\end{equation}
where $p$ is the slope of the power law accelerated electron
distribution, $\sigma_T$ is the Thomson cross section, $m_e$ is the electron
mass, $c$ is the speed of light, $q_e$ is the electron charge, $n'$ is the
comoving number density and $B'$ is the comoving magnetic field
strength. The field strength $B'$ is determined from the comoving
internal energy density $e'_i$ using $B' = \sqrt{ \epsilon_B e'_i 8
  \pi}$. Here $\epsilon_B$ is a free parameter that determines how
much energy is converted into magnetic energy near and behind the shock
front. The shape of the spectrum is determined by the synchrotron
critical frequency $\nu'_m$ and the cooling frequency $\nu'_c$. These
frequencies are set according to
\begin{equation}
 \nu'_m = \frac{3}{16} \left( \frac{p-2}{p-1} \frac{\epsilon_e e'_i}{n' m_e c^2}
\right)^2 \frac{q_e B'}{m_e c},
\end{equation}
where $\epsilon_e$ parameterizes the fraction of the internal energy density in
the shock-accelerated electrons, and
\begin{equation}
 \nu'_c = \frac{3}{16} \left( \frac{3 m_e c}{4 \sigma_T \epsilon_B e'_i t /
\gamma} \right)^2 \frac{q_e B'}{m_e c}.
\end{equation}
The cooling break is therefore estimated by using the duration of the
explosion as a measure of the cooling time: $t$ is the lab-frame time
and $\gamma$ is the Lorentz factor of the fluid element in the lab frame
(i.e. the frame of the unshocked medium) used to translate to the time
comoving with the fluid element. The emitted power at a given
frequency depends on the position of that frequency in the spectrum
with respect to $\nu'_m$ and $\nu'_c$ and the relative position of
$\nu'_m$ and $\nu'_c$ (\emph{fast cooling} when $\nu'_c < \nu'_m$ and
\emph{slow cooling} when $\nu'_m < \nu'_c$). The spectrum consists of
connected power law regimes. For example, for a fluid element the
emitted power at observer frequency $\nu$ ($\nu'$ in the comoving
frame) between synchrotron break and cooling in the slow cooling case
is given by $P' ( \nu' / \nu'_m )^{(1-p)/2}$. The complete shape of
the spectrum is given in \citet{Sari1998} and can also be found in the
Appendix. The power in the observer frame is then obtained by applying
the appropriate beaming factors and Doppler shifts to power and
frequency. Finally the received flux is calculated by taking into
account the luminosity distance and redshift (the latter is also used
to translate between observer frequency and comoving frequency).

In our calculations we have set $p = 2.5$, $\epsilon_e = \epsilon_B =
0.1$.  These are typical values found for afterglows. Different
redshifts have been calculated, but in this paper we only present
results where we have ignored redshift (i.e. $z \equiv 0$) and set the
observer luminosity distance at $d_L = 10^{28}$ cm. If the redshift is
increased the features of the light curves stretch out to later
observer times. Light curves computed for a range of observer
redshifts will be presented in an upcoming publication.

The main radiation results from ZM09 for an on-axis observer are the
following.  The jet break due to lateral expansion was found to be
weaker than analytically argued, while the jet break due to jet edges
becoming visible was stronger than expected from the simplest
analytical models, although not unexpected from calculations taking
limb-brightening into account. The weaker jet break due to lateral
spreading can be understood from the lateral spreading being
logarithmic instead of exponential as has been often assumed in
analytical models (see ZM09, Fig. 3). The long transition time to the
nonrelativistic regime for the blast wave was already mentioned in the
previous section. When the blast wave has become nonrelativistic the
counterjet is no longer beamed away from the observer. It becomes
distinctly visible around $t_{cj} = 2 ( 1+z) t_{NR}$, with the ratio
between flux from counter and forward jet peaking at 6 at 1 GHz at
3800 days for the simulation settings (at $z = 1$).

\section{Numerical Results}
\label{results_section}
\begin{figure*}
 \centering
 \includegraphics[width=\textwidth]{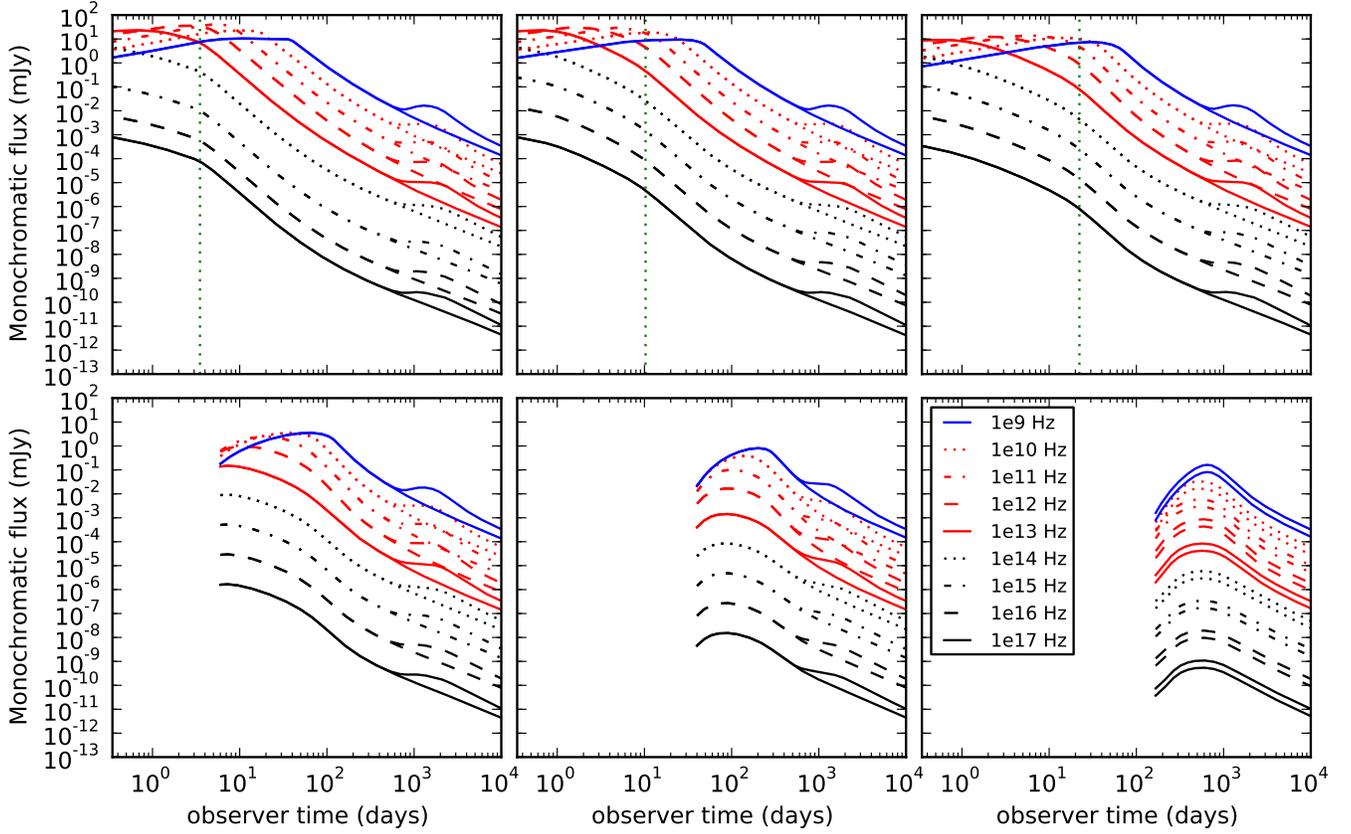}
 \caption{Simulated light curves for various observer angles and
   frequencies. On the top row we have small observer angles, 0.0, 0.1
   and 0.2 radians from left to right. On the bottom row we have large
   observer angles, 0.4, 0.8 and 1.57 radians from left to
   right. Observer frequencies from $10^9$ up to $10^{17}$ are
   plotted. For each frequency and angle a curve is plotted both with
   and without the contribution from the counterjet. The green
   vertical lines in the top plots indicate jet break time estimates
   using equation \ref{jetbreak_time_equation}. The legend in the
   bottom right plot refers to all plots. Large observer angles are
   truncated at earlier time to show only observation times completely
   covered by the simulation.}
\label{collective_figure}
\end{figure*}
\begin{figure}[h]
 \centering
 \includegraphics[width=1.0\columnwidth]{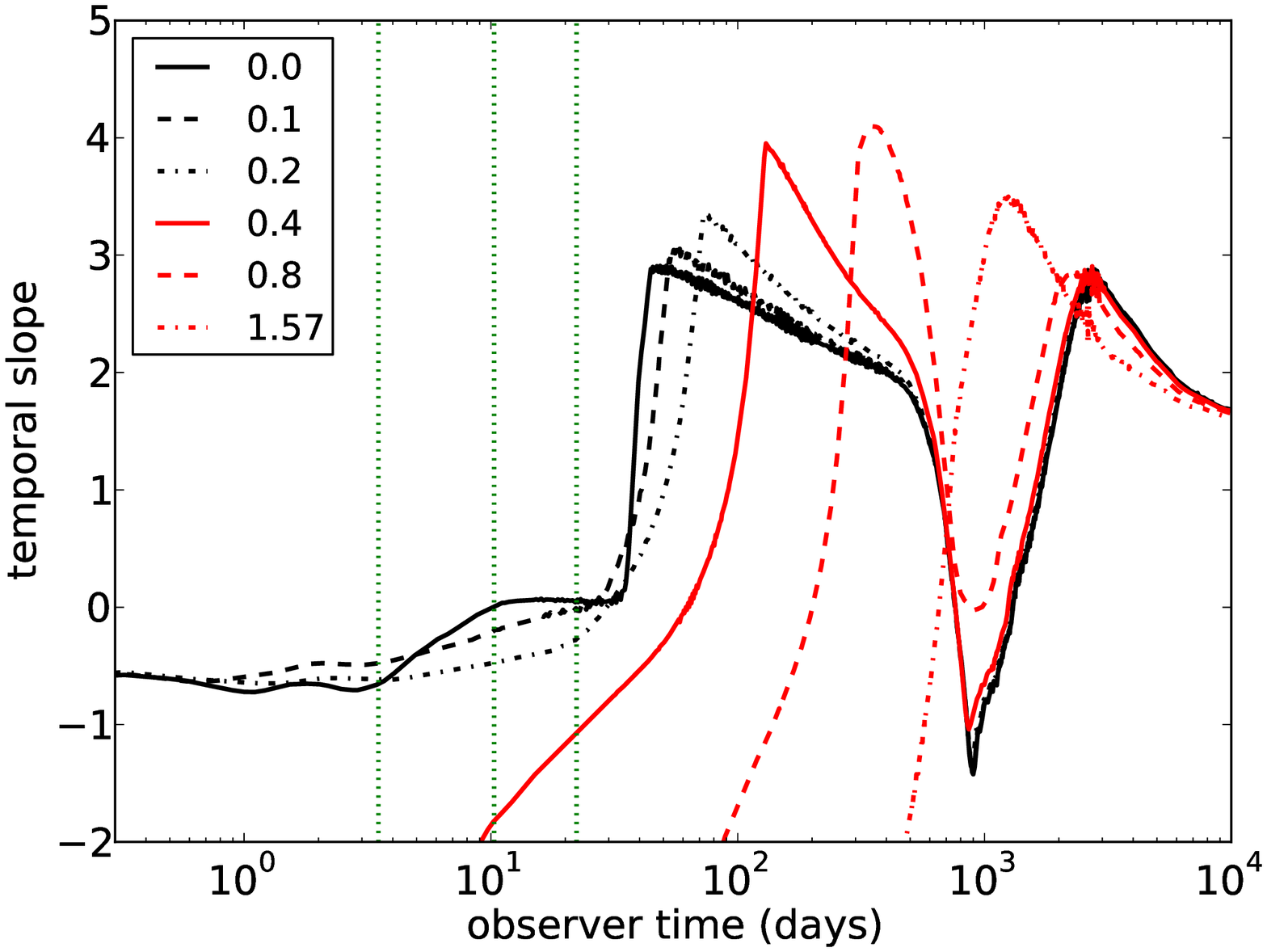}
 \caption{Temporal decay index $\alpha$ assuming $F \propto
   t^{-\alpha}$. The observer frequency is $10^9$ Hz. The vertical
   lines at 3.5 days indicate the jet break time estimates for
   observers at 0, 0.1 and 0.2 radians (from left to right), using
   equation \ref{jetbreak_time_equation}.  }
 \label{slopes_low_figure}
\end{figure}
\begin{figure}[h]
 \centering
 \includegraphics[width=1.0\columnwidth]{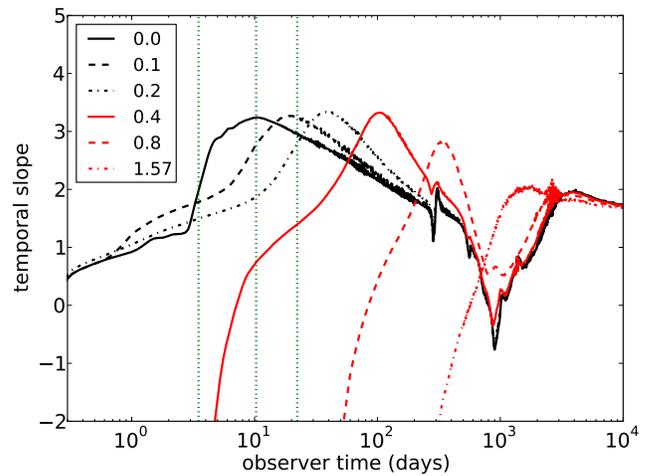}
 \caption{Temporal decay index $\alpha$ assuming $F \propto
   t^{-\alpha}$. The observer frequency is $10^{17}$ Hz. The vertical
   lines at 3.5 days indicate the jet break time estimates for
   observers at 0, 0.1 and 0.2 radians (from left to right), using
   equation \ref{jetbreak_time_equation}.}
 \label{slopes_high_figure}
\end{figure}
We first present the main results for both small and large observer
angles, looking both at the light curves and the corresponding
temporal slopes. An overview of light curves is shown in Fig.
\ref{collective_figure}, where we have plotted multi-frequency light
curves spanning from $10^9$ to $10^{17}$ Hz. The temporal slopes for
the lowest frequency $10^9$ Hz and the highest frequency $10^{17}$ Hz
are separately plotted in Figs. \ref{slopes_low_figure} and
\ref{slopes_high_figure}. For the on-axis results we estimate the jet
break, a combination of lateral spreading and jet edges becoming visible,
to occur around 3.5 days (the 7 days mentioned in ZM09 is for $z =
1$). Direct comparisons at the same frequency of different observer
angles are shown in figures \ref{model2_figure} and
\ref{supernovae_figure}, for 8.46 GHz.  As the observer angle
increases, the jet break splits into two for observers still inside
the jet. Once the observer is positioned at the jet edge only one
break remains that is significantly postponed. This effect is similar
across all frequencies. The steepest drop in slope is the one
associated with the edge of the jet furthest from the observer and can
therefore be estimated to occur around
\begin{equation}
t_j = 3.5 (1+z) E_{iso,53}^{1/3} n_1^{-1/3} \left(\frac{\theta_0 +
\theta_{obs}}{0.2}\right)^{8/3} \textrm{ days},
\label{jetbreak_time_equation}
\end{equation}

where $E_{iso,53}$ is the isotropic equivalent energy in units of
$10^{53}\,\erg$, $n_1$ is density of the medium in units of
$\cm^{-3}$, $\theta_0$ is the jet half opening angle, and
$\theta_{obs}$ is the observer angle relative to the jet axis.
Jet breaks can be used to estimate the opening angles of GRB jets.  It
is usually assumed in GRB afterglow modeling that the observer is on
the jet axis.  However, if the observer is near the edge of the jet, the
jet opening can be overestimated by a factor of up to 2, and the
beaming-corrected total energy can be overestimated by a factor of up
to 4.  For a typical observer at $\theta_{obs} \approx 2\theta_0/3 $,
the beaming-corrected energy can be overestimated by a factor of $\sim
3$.  The observational implications of this effect will be further
discussed in Section~\ref{discussion_section}.

At high observer angles, the rise of the light curve is postponed
until the point where relativistic beaming has weakened sufficiently
for the observer to be in the light cone of the radiating fluid.  Due
to limb-brightening the drop in temporal slope following a jet break
initially overshoots its asymptotic value. After that it starts to
change again due to the onset of the transition into the
nonrelativistic regime and the rise of flux from the counterjet,
before it finally settles into its asymptotic value for the
nonrelativistic regime. In order to put the simulation results in
context and to differentiate between the break due to lateral
spreading and the break due to the edges becoming visible we will
compare the simulation results against the BM solution for a hard
edged jet without lateral spreading in the next subsection.

\subsection{Afterglow emission for on-axis observer -- hydrodynamic
  simulation versus analytic model} 
\begin{figure}[h]
 \centering
 \includegraphics[width=1.0\columnwidth]{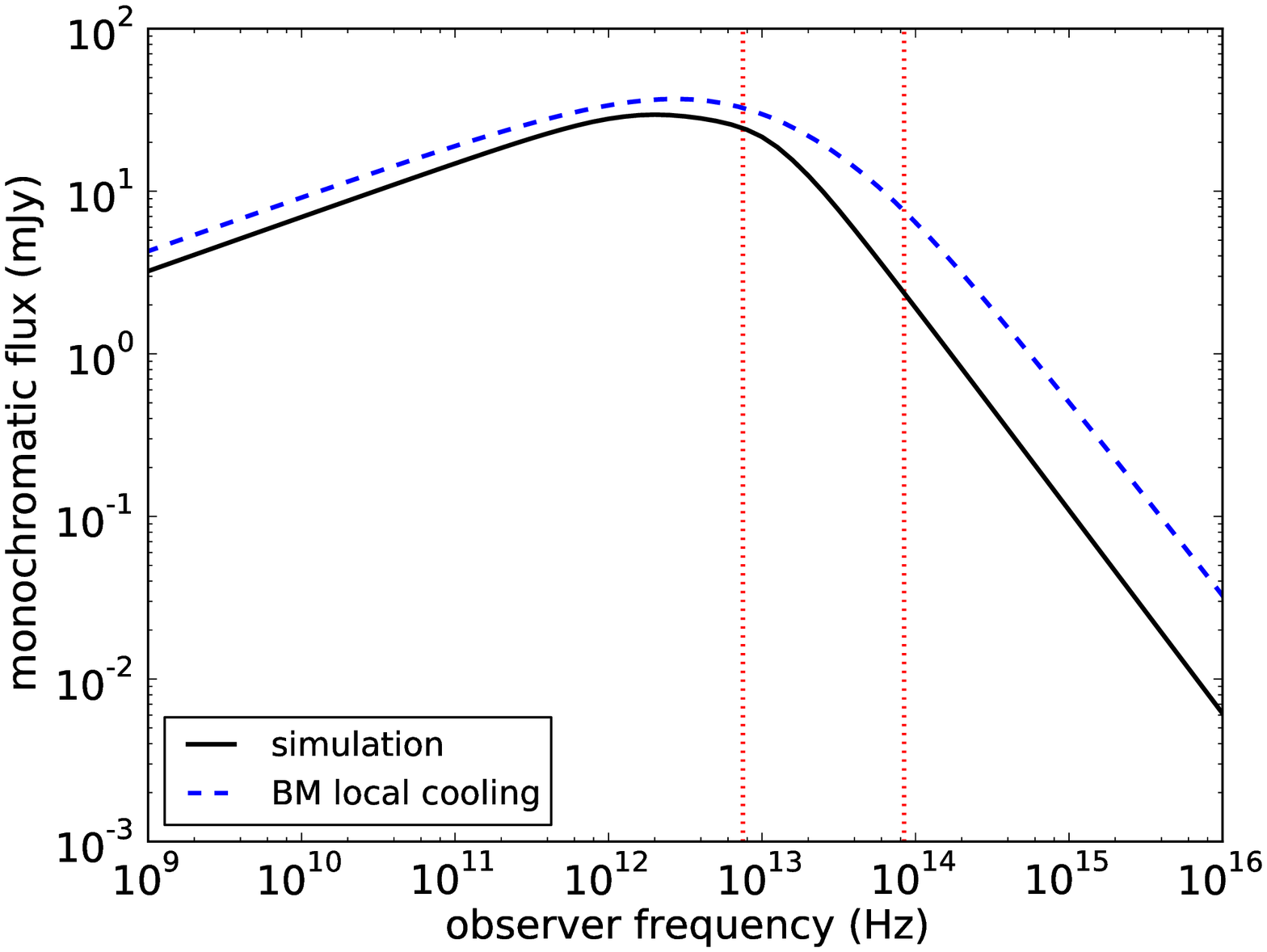}
 \caption{direct comparison between on-axis spectrum from simulation
   (solid curve) and BM solution with locally calculated cooling times
   (dashed curve), omitting self-absorption. The spectra are taken at
   1 day in observer time, so well before the jet break. The leftmost
   vertical dotted line denotes the analytically calculated position
   of $\nu_m$, the rightmost that of $\nu_c$ (calculated for the BM
   solution).}
 \label{exactspectrum_figure}
\end{figure}
In Fig. \ref{exactspectrum_figure} we show a comparison between
on-axis spectra at 1 day in observer time, calculated from the
simulation and from an analytical description using the BM solution
plus synchrotron emission \citep{vanEerten2009}. The observed time is
well before the jet break and before significant lateral spreading or
slowing down of the jet has occurred. With the dynamics for both the
simulation and the BM solution still being nearly equal, the figure
therefore mainly shows the difference between the two approaches to
synchrotron radiation. The differences below the cooling break $\nu_c$
are marginal and can be attributed to the absolute scaling of the
emitted power. The difference beyond $\nu_c$ is significantly
larger. The reason for this is that the simulation follows the
approach to electron cooling from \citet{Sari1998}, where the cooling
time is globally estimated by setting it equal to the duration of the
explosion, whereas \citet{vanEerten2009} build upon \citet{Granot2002}
and calculate the local cooling time for each fluid element, which is
given by the time passed since the fluid element has crossed the front
of the shock. When the cooling time is calculated locally, the
transition between pre- and post-cooling is smooth, with areas of the
fluid further downstream making the transition before those at the
front. Having the cooling time set globally results in a sudden global
transition between the pre- and post-cooling regimes instead, with the
same asymptotic spectral slope but a different value for the cooling
break frequency and therefore a different absolute scaling of the flux
for $\nu > \nu_c$. Note that even for a global cooling time, the
sudden transition in the emission frame will still get smeared out in
the observer frame.

\begin{figure}[h]
 \centering
 \includegraphics[width=1.0\columnwidth]{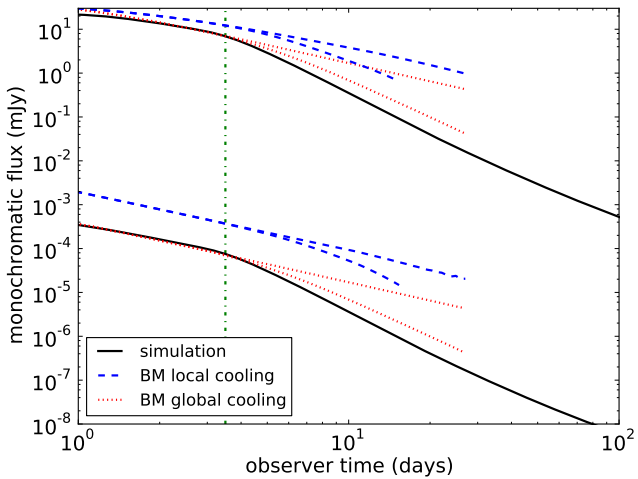}
 \caption{Direct comparison of simulation results and heuristic
   description based on the Blandford-McKee exact solution. Simulated
   light curves are shown for observer frequencies $10^{13}$ and
   $10^{17}$ Hz (top and bottom solid line respectively). Analytical
   light curves are drawn for global ({\it dashed}) and local ({\it
     dotted}) cooling respectively. The top line of each pair is for a
   spherical model whereas the lower line is for a conic section only.
   The 3.5 days estimate for the jet break is indicated with a
   vertical dotted line.}
 \label{exactlightcurve_figure}
\end{figure}
Figure \ref{exactlightcurve_figure} shows a direct comparison between
the on-axis light curves obtained from the simulation and light curves
from the same analytical model as before. If no lateral spreading is
assumed, a conic section of the spherically symmetric BM solution can
be used to show the difference between the jet break due to the edges
becoming visible and the combination of this break and the jet break
due to lateral spreading. The exact light curves shown in
Fig. \ref{exactlightcurve_figure} do not cover the entire observer
time span because the BM solution ceases to be valid around a shock
Lorentz factor of $\gamma \backsim 2$ (and slightly earlier for a
local cooling calculation). The light curves are truncated at the
point where this would start to affect the observed emission.

From the figure it can be seen that including lateral spreading has
the effect that the jet break becomes steeper and starts slightly
earlier (which confirms the low resolution comparison shown in
\citealt{vanEerten2010b}). The figure also shows the strong overshoot
in steepening of the light curve following the jet break, also seen in
Fig. \ref{slopes_high_figure}. This overshoot has also been discussed
in \citet{Granot2001} and ZM09.

The difference between the detailed local treatment of electron
cooling and the global treatment of cooling is important when applying
simulation results to actual data: it should be kept in mind that the
simulation light curves systematically underestimate the flux beyond
the cooling break. Electron cooling aside, the long term qualitative
behavior of the light curve is fully captured by the simulation, with
the results covering not only the relativistic regime but also the
non-relativistic regime and the transition in between.

\subsection{Afterglow emission for off-axis observer -- hydrodynamic
  simulation versus analytic model} 
\label{off_axis_comparison_section}

No exact solution exists that fully includes lateral spreading of the
jet. In the Appendix we describe a simplified analytical model that
approximates the behavior of the jet and allows us to calculate the
observed flux for an observer at an arbitrary angle. Many such models
exist in the literature (see e.g. \citet{Oren2004, Waxman2004,
  Soderberg2006, Huang2007} etc.) and our model is not strongly
different. Its distinguishing features are that it smoothly connects
the relativistic BM solution to the nonrelativistic ST solution and
that a conservative approach to lateral spreading is used where the
jets start to spread at the speed of sound (and therefore
logarithmically) upon approaching the nonrelativistic regime.
\begin{figure}[h]
 \centering
 \includegraphics[width=1.0\columnwidth]{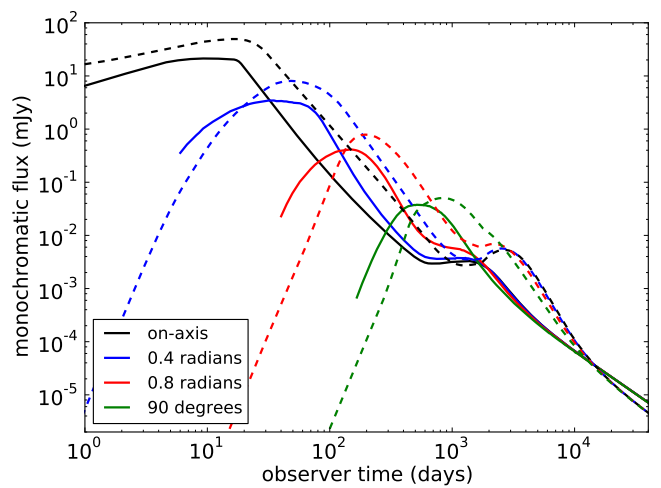}
 \caption{Direct comparison between simulation results (solid lines)
   and analytical model (dashed lines) for different opening angles,
   for radio frequency 8.46 Ghz.}
 \label{model2_figure}
\end{figure}

\begin{figure}[h]
 \centering
 \includegraphics[width=1.0\columnwidth]{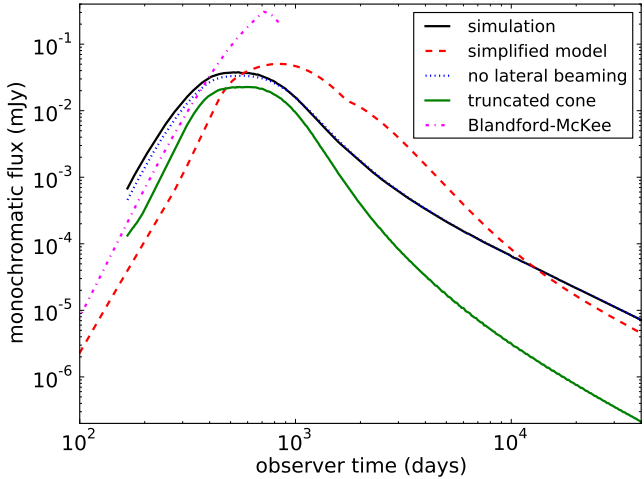}
 \caption{Light curves at 8.46 Ghz for an observer at $\theta_{obs} =
   90^\circ$.  The \emph{simulation} and \emph{simplified model}
   curves are repeated from fig.  \ref{model2_figure}. If we ignore
   the fluid velocity in the lateral direction for the purpose of the
   radiation calculation, we get the slightly lower curve labeled
   \emph{no lateral beaming}. The emission is still well in excess of
   a light curve from a radiation calculation that takes the BM
   profile instead of the simulated fluid profile as input, labeled
   \emph{Blandford-McKee}. Only by completely omitting the emission
   contribution from matter that has spread sideways out of the
   original jet opening angle, we get a flux level that is initially
   comparable to that of the exact BM solution and the simplified
   model.  This is shown by the \emph{truncated cone} curve.}
 \label{beaming_figure}
\end{figure}

In figure \ref{model2_figure} we show a comparison between off-axis
light curves generated using the analytical model and light curves
calculated from the simulation. Qualitatively both simulation and
model show the same features.  Quantitatively however, the differences
are substantial. The simulation light curves peak earlier than the
model light curves and do so at lower peak luminosity. At early times,
the emission from off-axis simulation light curves is higher than that
from the corresponding model light curves. The early time slopes for
simulation and model light curves are similar. However, the further
off-axis the observer, the later the observer time at which the
simulation provides full coverage. In the figure, we have truncated
the light curves at the observer time before which radiation from the
blast wave with $\gamma > 20$ would have been required. For two blast
wave jets viewed sideways ($\theta_{obs} = 90^\circ$), for example,
full coverage starts beyond 100 days. This means that, even though the
initial slopes agree between simulation and model, the early time
shape of off-axis light curves in reality will be largely dictated by
the initial shape of the blast wave, which does not need to be
anything like the BM solution. Collapsar jet simulations indicate the
existence of a cocoon around the emerging jet (see e.g.,
\citealt{Zhang_WM_2003_ApJ, Zhang_WH_2004_ApJ, 2007ApJ...665..569M,
  Mizuta2009}).

We can understand why the off-axis light curves from the simulation
are initially brighter than those from the simplified model by looking
at one of the angles in more details. In Fig. \ref{beaming_figure} we
have again plotted the simulation and model light curves for an
observer at $\theta_{obs} = 90^\circ$ (1.57 rad), together with a
number of variations. We have now also included a light curve where we
continue using the BM solution to determine the local fluid
conditions, instead of the dynamical simulation results, but otherwise
proceed as if we were reading the fluid quantities from disc (because
of this, the curve also serves as a consistency check on the radiation
calculation itself). The same approach has been used to generate the \emph{BM global cooling} light curve in Fig. \ref{exactlightcurve_figure}. The curve initially lies significantly below the
simulation light curve. It also lies above the simplified model
curve. The flux level of this BM curve is determined in part by the
numerical resolution that we assume. In the plot we have used a
resolution similar to that for the simulation curve, which initially
resolves the radial profile with approximately 17 cells. Increasing
the resolution moves the BM curve closer to the simplified slab curve,
and not to the simulation curve. The difference between the BM curve
and the simulation curve is real and we have added to the plot two
hybrid simulation / model curves to make clear the cause of this
difference.

First, when we completely ignore the velocity $v_\theta$ in the
angular direction for the purpose of calculating the emission but
otherwise still use the simulation dynamics, we find that the
resulting light curve, labeled \emph{no lateral beaming} in the
figure, initially lies somewhat below the simulation curve before the
two eventually merge. This tells us that part of the observed flux
level is caused by beaming towards the observer of material spreading
sideways, but that this is not the main cause of the difference
between simulation and the hard edge jet models. At late times beaming
no longer plays any role and the two curves are indeed no longer
expected to be different.

The main reason for the difference is shown by the second additional
curve.  When calculating the light curve labeled \emph{truncated cone}
in the figure we have omitted the contribution to the radiation of any
material that has spread sideways outside the original jet opening
angle. The resulting curve lies very close to the BM light curve at
first, before becoming orders of magnitude lower than all other
curves. The late time behavior is as expected, for then only a small
fraction of the energy and particle density is still contained within
the original opening angle; the actual simulation flow has become
roughly spherical. The early time behavior and the similarity between
the truncated cone and BM light curves is more relevant. It
demonstrates that the light curve for an off-axis observer is
dominated by the emission from material that has spread sideways out
of the original jet opening angle even though the energy of the
material is very little and the sideways spreading is not yet
dynamically important. In hindsight, the fact that the material on the
side of the jet dominates the observed radio flux can easily be
understood.  It is not so much due to the fact that it moves a little
faster towards the observer, for the $v_\theta$ component to the
beaming is not that strong (as we have shown above). It is instead due
to the fact that the radial velocity component $v_r$ drops quickly
outside of the original jet opening angle and as a result the material
outside the original jet opening angle is not beamed away from the
observer as much as the material in the original jet cone. By
contrast, for an on-axis observer the opposite is true and for a long
time the received flux is dominated by emission from material inside
the original jet opening angle. This has been demonstrated explicitly
by \citet{vanEerten2010b}.

In the above sections we did not discuss the effects of synchrotron
self-absorption. We will postpone addressing synchrotron
self-absorption, which is not included in the simulation, until
section \ref{orphan_afterglows_section}.

\section{Application: Hidden Jet Breaks?}
\label{jetbreaks_section}
A large number of X-ray afterglow light curves have been obtained by
the \emph{Swift} satellite since it was launched in 2004
\citep{Gehrels2004}. In a surprisingly large number of cases, these
light curves fail to show a clearly discernable jet break
\citep{Racusin2009, Evans2009}. Using our simulation results as a
basis to generate synthetic \emph{Swift} data sets for observers
positioned at different angles from the jet axis, we show that the
effect of observer position on the temporal evolution inferred from
the data can be profound and sometimes render the jet break difficult
to detect.

\subsection{Procedure for creating synthetic data}
The synthetic data sets that we produce should be comparable to those
produced by the on-line \emph{Swift} repository
\citep{Evans2007}. Also they should have data points at a sufficiently
late time that, if this were actual data, the jet break would be
considered missing and not merely delayed. We therefore make sure that
we have data up to at least 10 days, in accordance with the criteria
for their `complete' sample set by \citet{Racusin2009}. The observed
on-axis jet break for our simulation occurs roughly around three
days. Synthetic light curves have also been created from an underlying
model by \citet{Curran2008} (who find that even broken power law
models observed on-axis can occasionally be mistaken for a single
power law decline) and we follow the same procedure as described in
that paper, changing only the time span and adding an additional late
time data point if necessary. We then have:
\begin{itemize}
\item Constant counts and 1 $\sigma$ fractional error of 0.25 per data
  point.
 \item 94 minute orbits (47 min on/off due to \emph{Swift}'s low-Earth
   orbit).
 \item Fractional exposure drops from 1.0 to 0.1 after one day (when
   \emph{Swift} is usually no longer dedicated completely to observing
   the burst).
 \item Rate cut off at $5 \times 10^{-4}$ cts/s.
 \item Observed number of cts/s is scaled to 0.1 at 1.0 day.
\end{itemize}
\begin{figure}[h]
 \centering
 \includegraphics[width=1.0\columnwidth]{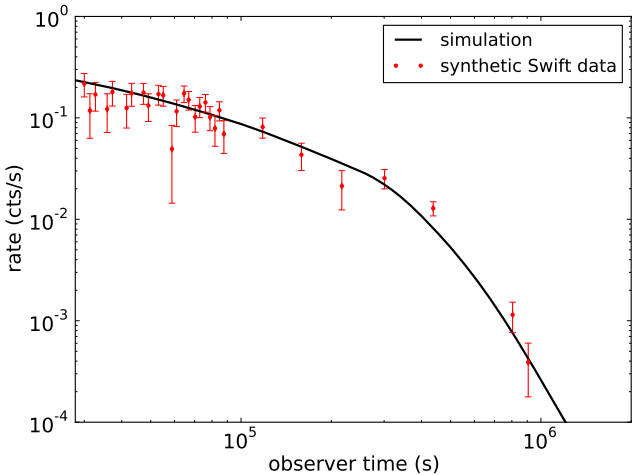}
 \caption{Synthetic Swift data generated from simulation light curve
   for an on-axis observer and $p = 2.5$. The simulation flux has been
   scaled to a count rate of 0.1 cts/s at 1 day observer time.}
 \label{synswiftonaxis_figure}
\end{figure}
\begin{figure}[h]
 \centering
 \includegraphics[width=1.0\columnwidth]{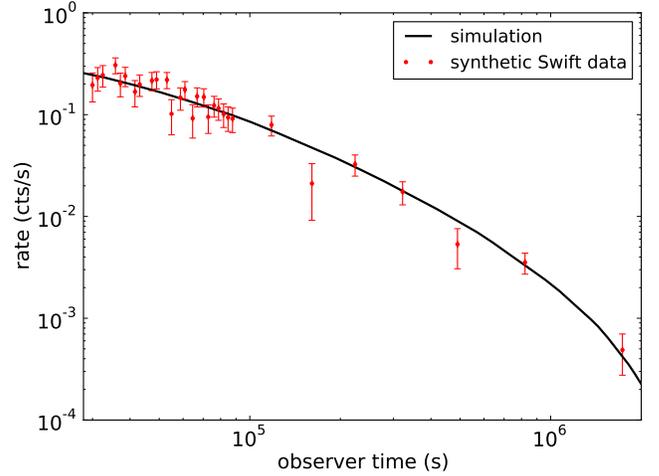}
 \caption{Synthetic Swift data generated from simulation light curve
   for an on-edge observer (i.e. at 0.2 rad) and $p = 2.5$. The
   simulation flux has been scaled to a count rate of 0.1 cts/s at 1
   day observer time.}
 \label{synswiftonedge_figure}
\end{figure}
We start generating data points from $3 \times 10^4$ s and continue
until the rate drops below $5 \times 10^{-4}$ cts/s. This starting
point is chosen such that we have full coverage from the simulation at
all observer angles under consideration. Like \citet{Curran2008}, we
increase the number of counts per bin to a number well in excess of
the numbers mentioned by \citet{Evans2007}. This has no physical
significance but is used to generate synthetic light curves containing
around 30 data points (The synthetic curves from Curran et al. contain
more data points because they use an earlier starting time). If needed
a late time data point is added to ensure that at least one data point
is observed after ten days. The last one or two data points, where the
count rates are less than $10^{-3}$ cts/s get a larger fractional
error of 0.5. Out of the thousands of synthetic curves that have been
generated, two randomly selected example synthetic light curves are
shown in figures \ref{synswiftonaxis_figure} and
\ref{synswiftonedge_figure}.

We generate light curves for observer angles 0.00, 0.02, 0.04, 0.06,
0.08, 0.1, 0.12, 0.14, 0.16, 0.18 and 0.2 radians. Note that by fixing
the count rate at 0.1 cts/s after one day, we end up comparing on-axis
observations to off-axis observations that are relatively brighter
(i.e. corresponding to closer GRBs). As can be seen in
Fig. \ref{collective_figure}, off-axis light curves are less bright
than on-axis curves for the same physical parameters.

\subsection{Fitting procedure and results}
We follow \citet{Curran2008} again when fitting the synthetic data
automatically using the \emph{simulated annealing} method to minimise
the $\chi^2$ of the residuals. The data are first fit to a single
power law, then to a sharply broken power law,
\begin{equation}
 F(t) = N \left\{ \begin{array}{rl} (t / t_b)^{-\alpha_1} & \text{if } t < t_b,
\\ (t / t_b)^{-\alpha_2} & \text{if } t > t_b. \end{array} \right.
\end{equation}
Curran et al. use a smooth power law, but a sharply broken power law
is also used by \citet{Racusin2009}. After each fit the count rates of
the data points are re-perturbed from their original on-model values
and a Monte Carlo analysis using 1000 trials is used to obtain average
values and $1 \sigma$ Gaussian deviations of the best fit parameters
and F-test probabilities. This process is repeated for the list of
observer angles mentioned previously. We have set $p = 2.5$.

The F-test is a measure of the probability $F_{prob}$ that the
decrease in $\chi^2$ associated with the addition of the two extra
parameters of the broken power law, $\alpha_2$ and $t_b$, is by chance
or not. When $F_{prob} \gtrsim 10^{-2}$ a single power law is commonly
favored, when $10^{-5} \lesssim F_{prob} \lesssim 10^{-2}$ neither is
favored but a single power law is usually presumed as the simpler
model and when $F_{prob} \lesssim 10^{-5}$ a broken power law is
favored.

\begin{figure}[h]
 \centering
 \includegraphics[width=1.0\columnwidth]{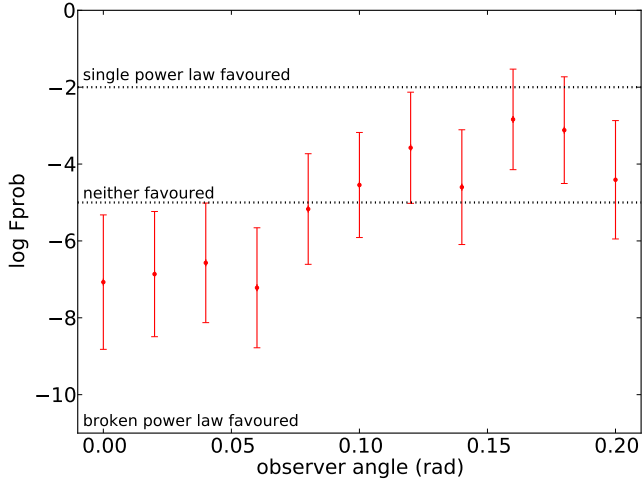}
 \caption{Results for F-test as function of observer angle. Note that
   a typical observer angle is 0.13.}
 \label{Fproblog_figure}
\end{figure}
\begin{table}
\centering
\begin{tabular}{l|llllll}
$\theta_{obs}$ & $\alpha$ & $\alpha_1$ & $\alpha_2$ & $t_b$ ($10^5$ s)\\
\hline
 0.00 & $1.69 \pm 0.040$ & $0.81 \pm 0.16$ & $3.32 \pm 0.79$ & $2.3 \pm 0.80$ \\
 0.02 & $1.68 \pm 0.037$ & $0.76 \pm 0.17$ & $3.14 \pm 0.73$ & $2.1 \pm 0.83$ \\
 0.04 & $1.63 \pm 0.038$ & $0.81 \pm 0.15$ & $2.96 \pm 0.61$ & $2.1 \pm 0.72$ \\
 0.06 & $1.78 \pm 0.036$ & $0.84 \pm 0.16$ & $3.05 \pm 0.58$ & $2.1 \pm 0.70$ \\
 0.08 & $1.52 \pm 0.040$ & $0.78 \pm 0.19$ & $2.26 \pm 0.37$ & $1.4 \pm 0.54$ \\
 0.10 & $1.51 \pm 0.040$ & $0.80 \pm 0.20$ & $2.07 \pm 0.26$ & $1.2 \pm 0.45$ \\
 0.12 & $1.49 \pm 0.042$ & $0.91 \pm 0.26$ & $1.95 \pm 0.29$ & $1.2 \pm 0.55$ \\
 0.14 & $1.65 \pm 0.035$ & $1.07 \pm 0.16$ & $2.36 \pm 0.52$ & $2.0 \pm 1.1$ \\
 0.16 & $1.46 \pm 0.042$ & $1.00 \pm 0.28$ & $1.95 \pm 0.47$ & $1.6 \pm 0.89$ \\
 0.18 & $1.40 \pm 0.040$ & $0.91 \pm 0.36$ & $1.85 \pm 0.50$ & $1.6 \pm 0.97$ \\
 0.20 & $1.42 \pm 0.036$ & $0.86 \pm 0.21$ & $2.01 \pm 0.45$ & $1.9 \pm 1.0$ \\
\end{tabular}
\caption{Average temporal power law slopes and jet break times $t_b$
  for 1000 Monte Carlo iterations per observer angle
  $\theta_{obs}$. Here $\alpha$ denotes the single power law slope,
  $\alpha_1$ the pre-break broken power law slope and $\alpha_2$ the
  post-break broken power law slope.}
\label{fit_results_table}
\end{table}
\begin{table}
\centering
\begin{tabular}{l|llllll}
$\theta_{obs}$ & $\log( F_{prob} )$ & single & ambiguous & broken \\
\hline
 0.00 & $-7.1 \pm 1.7$ & 0 & 130 & 870 \\
 0.02 & $-6.9 \pm 1.6$ & 0 & 148 & 852 \\
 0.04 & $-6.6 \pm 1.6$ & 0 & 202 & 798 \\
 0.06 & $-7.2 \pm 1.6$ & 0 & 60 & 940 \\
 0.08 & $-5.2 \pm 1.4$ & 6 & 510 & 484 \\
 0.10 & $-4.5 \pm 1.4$ & 22 & 698 & 280 \\
 0.12 & $-3.6 \pm 1.4$ & 129 & 767 & 104 \\
 0.14 & $-4.6 \pm 1.5$ & 21 & 641 & 338 \\
 0.16 & $-2.8 \pm 1.3$ & 274 & 675 & 51 \\
 0.18 & $-3.1 \pm 1.4$ & 259 & 686 & 55 \\
 0.20 & $-4.4 \pm 1.5$ & 37 & 663 & 300 \\
\end{tabular}
\caption{Average $\log(F_{prob})$ values and classifications based on
  the criteria described in the text. For each angle, the
  classifications add up to 1000 Monte Carlo runs. }
\label{break_times_results_table}
\end{table}
The results of the fitting procedure are shown in tables
\ref{fit_results_table}, \ref{break_times_results_table} and fig.
\ref{Fproblog_figure}. In table \ref{fit_results_table} any slope
$\alpha$ and its error $\Delta \alpha$ means that if an observer fits
data to a swift dataset with similar count rate and duration and if
the swift data was produced by an explosion that is accurately
described by our numerical model, then that observer would find a
slope that lies within $\Delta \alpha$ of $\alpha$ (the errors
therefore do not reflect on the accuracy of the Monte Carlo run, which
has converged to far greater precision).  The analytically expected
slopes for an on-axis observer and $p=2.5$ are 1.375 before and 2.125
after the jet break. These values are not even reproduced within $1
\sigma$, which for the post-break slope can largely be attributed to
the overshoot in slope directly after the break (see also section
\ref{results_section}). Although the inaccuracy of the pre-break slope
inferred from (synthetic) light curves will be smaller when earlier
time data is available, which is often the case for \emph{Swift} light
curves, a systematic difference will remain (see also
\citealt{Johannesson2006}). This emphasizes the importance of
numerical modeling for the proper interpretation of \emph{Swift}
data. The situation gets worse when we move the observer off-axis. As
Fig. \ref{Fproblog_figure} shows, even though a jet break is clearly
detected on-axis for our physics settings, it can become hard to
distinguish from a single power law for an observer positioned at
$\theta_j / 2$. The average observer angle one would expect when
observing jets oriented randomly in the sky is $2\theta_j /3$ (for
small jet opening angles, and assuming that the observer angle lies
between zero and the jet opening angle). Larger observer angles will
lead to \emph{orphan afterglows} that we discuss separately below. It
therefore follows that a significant number of jet breaks may remain
hidden in the data due to the jets not being observed directly
on-axis. In table \ref{break_times_results_table} the classifications
for the individual Monte Carlo iterations are also counted. Assuming
again that jets are oriented randomly in the sky and that jets are
only observed up to observer angles equal to $\theta_j$, we can
calculate how often an afterglow with the physical parameters of the
simulation would be classified as showing a jet break. For each angle
$\theta_i$ we have classified $n_i$ out of 1000 synthetic curves as
showing a jet break. This means that we will classify the afterglow
described by the simulation as showing a jet break only $\sum_i n_i
\sin( \theta_i ) / \sum_i 1000 \sin( \theta_i ) \times 100 \% \approx
29 \%$ of the time. This value is only a very rough estimate, for it
depends on the $F_{prob} \lesssim 10^{-5}$ criterion, that is to some
extent arbitrary. Also, as noted before, the off-axis curves are
relatively brighter due to the fixed count rate. The requirement of
having data up to ten days introduces a selection effect as well.

For all results above, it should be kept in mind that they have been
obtained for a single half opening angle of 0.2 radians
(approx. $11.5^\circ$) which is relatively large (although not
extremely so and within the range of jet opening angles
observationally inferred from \emph{Swift} data). This results in a
later jet break than a smaller opening angle would lead to. On the
other hand we have set $z = 0$, which again moves the jet break to
earlier observer times and thereby compensates for the large opening
angle. The general effect of higher observer angles is that both jet
edges become observable at different times. The reason that a broken
power law did not always produce a significantly better fit than a
single power law is mainly because the full drop in count rate
associated with the further edge got pushed out beyond ten days for
off-axis observers (Section~\ref{results_section}). For strongly
collimated jets with small opening angles this effect is therefore
expected to be less severe.

The often large difference between the physical parameters like $p$
(affecting the slope of the light curve) and $\theta_j$ (affecting the
break time) used in a model and the values for these parameters when
inferred from synthetic data created from that model has been
discussed in detail by \citet{Johannesson2006}. They also include the
observer angle as a model parameter, but do not discuss it further in
their paper.

\section{Application: Orphan Afterglow Searches} 
\label{orphan_afterglows_section}

The existence of orphan afterglows is an important and general
prediction of current afterglow theories. Regardless of the GRB
launching mechanism and the initial baryon content of the jet,
eventually synchrotron emission from a decelerating baryonic blast
wave should be observable for any observer angle.  For this reason,
various groups have looked for orphan afterglows, both at the optical
and radio frequencies (e.g. \citealt{Levinson_etal_2002_ApJ,
  Gal-Yam_etal_2006_ApJ, Soderberg2006, Malacrino_etal_2007_AA}). Few
positive detections have been reported and surveys and archival
studies have mainly served to establish constraints on GRB rates and
beaming factors.  \citet{Soderberg2006}, for example, conclude from
late time radio observations of 68 local type Ibc supernovae (SNe)
that less than $\backsim 10 \%$ of such SNe are associated with GRBs,
and constrain the GRB beaming factor to be $\left< (1 -
  \cos{\theta_j})^{-1} \right> \lesssim 10^4$. A lower limit to the
beaming factor of $\left< (1 - \cos{\theta_j})^{-1} \right> \gtrsim
13$ is provided by \citet{Levinson_etal_2002_ApJ}.

\begin{figure}[h]
 \centering
 \includegraphics[width=1.0\columnwidth]{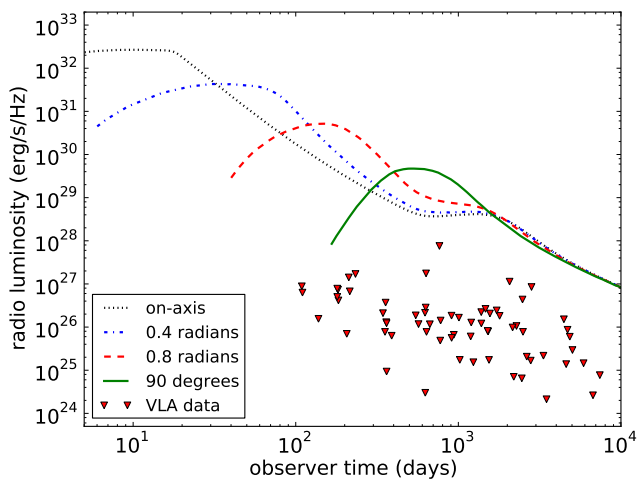}
 \caption{VLA late time radio limits ($3 \sigma$) for 66 local type
   Ibc of supernovae compared against simulation results. All
   supernova redshifts have been ignored (the largest redshift, that
   of SN 1991D, is $\backsim 0.04$). The fluxes have been rescaled to
   luminosities.  All VLA observations were done at 8.46 GHz, the
   afterglow light curve is calculated at the same frequency. As in
   the rest of this paper, the simulation jet half opening angle is
   $11.5^\circ$.}
 \label{supernovae_figure}
\end{figure}
Such estimates require a model describing the shape of off-axis
light-curves, and are therefore sensitive to model
assumptions. Eventually, comparing observations and detailed
simulations like the one described in this paper will place the most
accurate observational limits on orphan afterglow characteristics. A
large number of simulations is required to fully explore the afterglow
parameter space. We can however, use the single simulation of this
paper that has typical values for the explosion parameters to confirm
the result from \citet{Soderberg2006} that their sample of 68
supernovae observations are all significantly fainter than a standard
afterglow viewed off-axis. This confirmation is shown in
fig.~\ref{supernovae_figure}, where we have plotted 66 supernovae
radio upper limits (omitting SN 1984L and SN1954A, which were not
observed at 8.46 GHz, from the original 68) together with our off-axis
simulated light curves. Note that the jet half opening angle in our
simulation is $11.5^\circ$, whereas \citet{Soderberg2006} use
$5^\circ$. The fact that the early time flux received by an off-axis
observer is actually stronger than analytically expected (as shown in
section \ref{off_axis_comparison_section}, where model and simulation
are compared directly) only strengthens the case made by Soderberg et
al.

\begin{figure}[h]
 \centering
 \includegraphics[width=1.0\columnwidth]{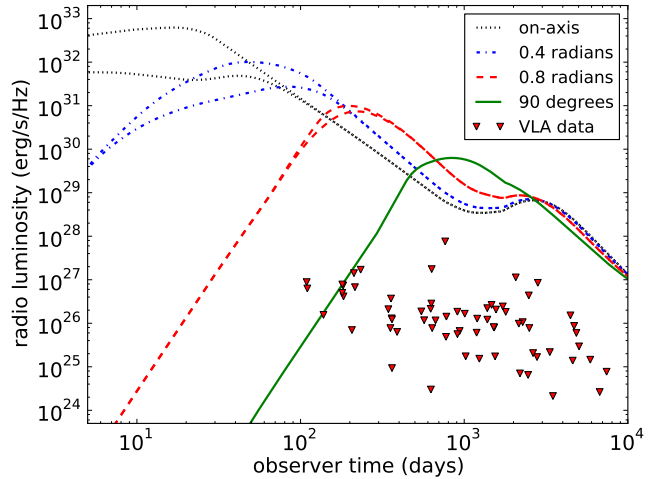}
 \caption{Analytically calculated light curves for different observer
   angles, with and without self-absorption. The observer frequency is
   set at 8.46 GHz.  Self-absorption influences the light curves only
   until a few hundred days at most. At 8.46 GHz, the light curve for
   an observer at 90 degrees is the same with and without
   self-absorption enabled. The jet half opening angle is
   $11.5^\circ$. The VLA late time radio limits for the local Ibc SNe
   are included as well.}
 \label{selfabsorption_figure}
\end{figure}
A possible caveat to the above is that our simulation light curves do
not include the effect of synchrotron self-absorption. Although we
cannot completely rule out that this plays a role without actually
calculating it, we can nevertheless look at the effect of
self-absorption on the model light curves, having already established
that model and simulation lead to at least qualitatively similar light
curves in section~\ref{off_axis_comparison_section}. In
Fig.~\ref{selfabsorption_figure} we show model light curves with and
without synchrotron self-absorption, calculated as explained in the
appendix. The figure shows that the effect of self-absorption is
initially significant for an on-axis observer but becomes less
pronounced for observers further off-axis. For an observer at
$90^\circ$, the light curves with and without self-absorption are
effectively identical. Aside from the minimal differences due to the
analytical model assumptions, the main difference between this figure
and fig. 1 from Soderberg et al. is due to the different jet opening
angles. For our wider jet opening angle, only the two earliest
supernovae lie clearly above the $90^\circ$ curve.

\section{Summary and Discussion}
\label{discussion_section}

In this paper we present broadband GRB afterglow light curves
calculated assuming synchrotron emission from a high-resolution
relativistic jet simulation in 2D. We have expanded the work presented
in \citet{Zhang2009} to include observers positioned off the jet
symmetry axis, both at small and large angles.  For the jet simulation
we have used the \textsc{ram} adaptive-mesh-refinement code, starting
from the Blandford-McKee analytical solution and letting the jet
evolve until it has reached the Sedov-Taylor stage and has
decollimated into a nearly spherical outflow. We have implemented
synchrotron radiation as described in \citet{Sari1998}.

When put in the context of analytical light curve estimates, our
simulations show the following:
\begin{itemize}
\item For an on-axis observer, the jet break from a 2D simulation
  including lateral spreading of the jet is seen earlier than that of
  a hard-edged jet. However, the break due to the jet edges becoming
  visible still dominates the shape of the light curve.
\item We compared a description of electron cooling that takes into
  account the local cooling time since the shocked electrons passed
  the shock front to a description that uses a global cooling time
  estimate (as we have done in the simulation). The latter approach
  underestimates the observed cooling break frequency and therefore
  the post-break flux as well.
\item The simulation light curves show that simplified homogeneous
  slab analytical models are qualitatively correct as long as they
  include a clear contribution from the counterjet for observers at
  all angles.
\item Contrary to what has thus far been assumed in simplified
  analytical models and even though lateral spreading of the jet is
  initially not dynamically important, the received flux for an
  off-axis observer is strongly dominated by emission from material
  that has spread laterally outside the original jet opening
  angle. This is due to the fact that material outside the original
  jet cone has slowed done considerably in the radial direction and is
  therefore not beamed away from the observer as much as material
  closer to the jet axis.
\item Moving the observer off-axis splits the jet break in two. The
  steep drop in slope only occurs after the break associated with the
  farthest edge and this break can thus be postponed until several
  weeks after the burst, even for observers positioned still within
  the jet cone. The late break time can be estimated by using the sum
  of the observer angle and the jet half opening angle instead of just
  the jet half opening angle. 
\end{itemize}

In addition to these direct numerical results, we have presented two
applications of our numerical work, one with observers positioned at
small and moderate observer angles and one with observers at large
observer angles.

\begin{itemize}
\item X-ray Light curves for observers at small observer angles are
  relevant for satellites such as \emph{Swift}. Recent authors
  (e.g. \citealt{Racusin2009, Evans2009}) have noted a lack of jet
  breaks visible in the data. In order to check whether the observer
  angle can cause the jet break to remain hidden in the data we have
  performed a Monte Carlo analysis where we created synthetic
  \emph{Swift} data out of simulation light curves for different
  observer angles. Observational biases and a Gaussian observational
  error were included. Broken and single power laws were then fit to
  the synthetic data sets. We found that it is not difficult to bury a
  jet break in the data for an off-axis observer. For our explosion
  parameters, even an observer at an angle of $\theta_j / 2$ will not
  find a significantly better fit for a broken power law than for a
  single power law, leading to a missing jet break. For a random
  observer angle within the cone of the jet, a synthetic light curve
  created from our simulation will only show a discernable jet break
  29 \% of the time. Although our simulation is somewhat atypical in
  its above average initial opening angle of the jet, these results
  nevertheless imply that the observer angle has a strong influence on
  the interpretation of X-ray data.  This holds even for observers
  still within the cone of the jet, observer angles that have usually
  been ignored and considered practically on-axis.

\item As a second application we have confirmed the result of
  \citet{Soderberg2006} that a sample of 68 nearby type Ibc SNe cannot
  harbor an off-axis GRB, at least for the typical afterglow
  parameters that we have used for our simulation. This confirms the
  observational restrictions placed on orphan afterglow rate and
  beaming factor by these and other authors. The fact that at early
  times the light curves for off-axis observers are actually brighter
  than analytically expected, even strengthens the conclusions of
  \citet{Soderberg2006}.
\end{itemize}

Recent deep late-time optical observations by \citet{Dai2008} have
detected jet breaks in several bursts and they suggest that the lack
of jet breaks in \emph{Swift} bursts is due to the lack of well
sampled light curves at late times.  Therefore they conclude that the
collimated outflow model GRBs is still valid.  However, the
non-detection of a jet break or a break at very late times in several
GRBs (050904, \citealt{Cenko2010}; 070125, \citealt{Cenko2010};
080319B, \citealt{Cenko2010}; 080721, \citealt{Starling2009}; 080916C,
\citealt{Greiner2009}; 090902B, \citealt{Pandey2010}; 090926A,
\citealt{Cenko2010b}) seems to infer a huge amount of released energy
($\gtrsim 10^{52}\,\mathrm{erg}$) in these bursts.  This leads
\citet{Cenko2010, Cenko2010b} to propose a class of
\emph{hyper-energetic} GRBs and challenge the magnetar model for GRBs
\citep{Usov92,Duncan92,Thompson04,Uzdensky07,Kom07,Buc09}.  Although
\emph{hyper-energetic} GRBs might exist, we propose an alternate
explanation, in which the observer is simply off-axis.  Note that an
observer is more likely to be off-axis than on-axis. A typical
observer sees the burst from $\theta_{obs} \approx 2\theta_0/3 $ so is
closer to the jet edge than jet axis. Applying an on-axis model to a
jet that is seen off-axis will \emph{overestimate} the total energy
release by a factor of up to 4 because the jet break can be delayed by
a factor of up to $\sim 6$ (see Eq. \ref{jetbreak_time_equation}).  We
emphasize that when the jet opening angle is corrected for off-axis
observer angle, the inferred energy release of those
\emph{hyper-energetic} events can be revised downwards by factors of
several lessening tension with magnetar models for the GRB central
engine.

These are the main conclusions of the work presented in this paper. In
addition to this, the current results also raise a number of issues
that need to be addressed in future work. Our simulation results can
be generalized by performing additional simulations.  And given the
quantitative differences between simulation and simplified analytical
models (such as the homogeneous slab model described in the Appendix),
this is expected to result in different constraints on orphan
afterglow rate and beaming factor. We have so far ignored
self-absorption when calculating light curves. A simplified
homogeneous slab approximation indicates that self-absorption is not
expected to play a large role for observers at high angles. The
applicability of our analytical model is however limited, especially
in view of our finding that the early flux received by an off-axis
observer is dominated by emission from material that has spread
sideways and has slowed down more than material on the jet axis. It is
conceivable that emission from this material has different spectral
properties. Finally, the significant difference in observed flux
between two approaches to synchrotron cooling that we have discussed
emphasize the importance of a detailed model for the microphysics and
radiation mechanisms involved.

\acknowledgments

We thank Peter A. Curran for allowing us the use of his computer code
for synthesizing and fitting Swift data and for helpful
discussion. This work was supported in part by NASA under Grant
No. 09-ATP09-0190 issued through the Astrophysics Theory Program
(ATP).  The software used in this work was in part developed by the
DOE-supported ASCI/Alliance Center for Astrophysical Thermonuclear
Flashes at the University of Chicago.
\appendix

\section{Summary of analytical model}
\label{summary_model_section}
Here we provide a short summary of our analytical model. In this
model, light curves are calculated by numerical integration over an
infinitesimally thin homogeneous blast wave front. The received flux
is given by
\begin{equation}
F = \frac{1}{4 \pi d^2_L} \int d^3 \mathbf{r}
\frac{\epsilon'_{\nu'}}{\gamma^2 (1-\beta \mu)^2}, 
\end{equation}
when we ignore the redshift $z$. Here $\epsilon'_{\nu'}$ is the
comoving frame emissivity and $\mu$ the angle between observer
direction and local velocity. The dependence of the beaming and
emissivity on the observer time $t_{obs}$ is kept implicit (see also
eqn. \ref{te2tobs_equation}; the volume integral needs to be taken
across different emission times).  Assuming the radiation is produced
by an infinitesimally thin shell with width $\Delta R$ we get
\begin{equation}
F = \frac{1}{4 \pi d^2_L} \int d \theta d \phi R^2 \sin{\theta} \Delta R
\frac{\epsilon'_{\nu'}}{\gamma^2 (1-\beta \mu)^2},
\label{flux_model2_equation} 
\end{equation}
For every observer time we integrate over jet angles $\theta$ and
$\phi$ (we define $\theta$ such that it is \emph{not} the angle
between observer and local fluid velocity, but that between local
fluid velocity and jet axis), while taking into account that radiation
from different emission angles.

The emissivity can be calculated from the local fluid conditions,
which we know in turn in terms of emission time $t_e$. For the blast
wave radius we have by definition:
\begin{equation}
 R = c \int \beta_{sh} d t_e,
\end{equation}
Where the subscript $sh$ indicates \emph{shock} velocity. From the
shock jump conditions it follows for arbitrary strong shocks that
\begin{equation}
 e'_{th} = ( \gamma - 1) n' m_p c^2.
\end{equation}
The comoving downstream number density $n'$ in both the relativistic
and nonrelativistic regime is given by
\begin{equation}
 n' = 4 n_0 \gamma,
\end{equation}
with $\gamma \to 1$ in the nonrelativistic limit. We assume this
equation to remain valid in the intermediate regime as well. This is
not implied by the expression above, where we have kept implicit the
dependence on the fluid adiabatic index (which changes from $4/3$ to
$5/3$ over the course of the blast wave evolution).

We set the width of the shell at a single emission time by demanding
that the shell contains all swept-up particles, leading to:
\begin{equation}
 4 \pi \left[ R_f (t_f) - R_b(t_f) \right] R^2 n = 4/3 \pi R^3 n_0 \to \left[
R_f (t_f) - R_b(t_f) \right] = \frac{R}{12 \gamma^2},
\end{equation}
where we have used $n = \gamma n' = 4 n_0 \gamma^2$, again assumed
valid throughout the entire evolution of the fluid. The subscript $f$
denotes the front of the shock and the subscript $b$ denotes the back
of the shock. Setting the shock width through the number of particles
is to some extent an arbitrary choice, and we could also have used the
total energy which would have yielded a different width (since the
downstream energy density profile is different from the downstream
number density profile). The width of the shell $\Delta R$ in equation
\ref{flux_model2_equation} has to take into account the emission time
difference between the front and back of the shell and is given by
\begin{equation}
 \Delta R = | R_f (t_f) - R_b (t_b) | = | R_f (t_f) - R_f( t_b ) \left[ 1 -
\frac{1}{12 \gamma(t_b)^2} \right] |.
\end{equation}
Because the shell is very thin, $R_f (t_b) \approx R_f (t_f) -
\beta_{sh} c \Delta t$. We integrate over emission arriving at a
single observer time, and for given values of $\mu$ and $t_{obs}$ we
have
\begin{equation}
 t_{obs} = t_f - \mu R_f (t_f) / c = t_b - \mu R_b ( t_b) / c,
\label{te2tobs_equation}
\end{equation}
which yields $\Delta R = \Delta t c / \mu$ when
differentiated. Combining the above, we eventually find
\begin{equation}
 \Delta R = \frac{1}{1 - \beta_{sh} \mu} \cdot \frac{R}{12 \gamma^2}.
\end{equation}

For the shock velocity we have
\begin{equation}
 (\beta_{sh} \gamma_{sh})_{BM} = \left( \frac{17 \cdot E_\mathrm{iso}}{8 \pi n_0
m_p c^5} \right)^{1/2} t_e^{-3/2} \equiv C_{BM} t_e^{-3/2}; \qquad (\beta_{sh}
\gamma_{sh})_{ST} = \frac{2}{5} \cdot 1.15 \cdot \left( \frac{E_j}{n_0 m_p c^5}
\right)^{1/5} \cdot t_e^{-3/5} \equiv C_{ST} \cdot t_e^{-3/5},
\end{equation}
in the BM and ST regime respectively. We artificially combine the two
simply by adding them (after squaring):
\begin{equation}
\beta_{sh}^2 \gamma_{sh}^2 = C^2_{BM} t_e^{-3} + C^2_{ST} t_e^{-6/5}.
\end{equation}
Note that the BM quantities depend on $E_{iso}$, while the ST
quantities depend on $E_{j}$. The two are related via $E_j = E_{iso}
\theta_j^2 / 2 $. Here $E_j$ is the total energy in \emph{both} jets,
and $\theta_j$ the \emph{half} opening angle of a jet. The fluid
Lorentz factor in the relativistic regime is related to the shock
Lorentz factor via $\gamma^2 = \gamma_{sh}^2 / 2$, while the fluid
velocity in the non-relativistic regime is related to the shock
velocity via $\beta = 3/4 \beta_{sh}$. We therefore construct a
relationship between emission time and fluid velocity similar to that
between emission time and shock velocity:
\begin{equation}
 \beta^2 \gamma^2 = \frac{1}{2} C_{BM}^2 t_e^{-3} + \frac{9}{16} C_{ST}^2
t_e^{-6/5}.
\end{equation}

We assume that the jet does not spread sideways throughout the
relativistic phase of its evolution. However, at some point the blast
wave \emph{must} become spherical. For if we keep the opening angle
fixed, but do take $E_j$ to dictate the ST solution instead of
$E_{iso}$ we would underestimate the final flux by integrating over an
integration domain that is too small. We will assume that the jet
starts spreading sideways when it has reached the nonrelativistic
phase and we take this moment to be given by \begin{equation} (\beta
  \gamma)_{BM} = 1 \to t_{NR} = 2^{1/3} C_{BM}^{2/3}.
\end{equation}
At this point, the jet starts spreading sideways with the speed of
sound $c_s$, leading to
\begin{equation}
R \frac{d \theta}{d t_{obs}} = c_s.
\end{equation}
In the nonrelativistic regime $d t_{obs}$ and $d t_e$ are
identical. The ST solution for the speed of sound for adiabatic index
$5/3$ is given by $c_s = r/\sqrt{20} t$, leading to
\begin{equation}
\theta = \theta_j + \frac{1}{\sqrt{20}} \ln \frac{t_{obs}}{t_{NR}},
\end{equation}
until spherical symmetry is reached. By contrast, Rhoads '99 take
$\Omega \approx \pi ( \theta_j + c_s t' / c t_e )^2$ (where $t'$ time
in the comoving frame, and $\Omega$ a solid angle) as the starting
point.

From the local fluid conditions the local emissivity can be
calculated. In the case of slow cooling, we define
\begin{eqnarray}
 \epsilon'_{\nu'} & = & \epsilon'_m \left( \frac{\nu'}{\nu'_m} \right)^{1/3},
\qquad \nu' < \nu'_m  \nonumber \\
 \epsilon'_{\nu'} & = & \epsilon'_m \left( \frac{\nu'}{\nu'_m}
\right)^{(1-p)/2}, \qquad \nu'_m < \nu' < \nu'_c \nonumber \\
 \epsilon'_{\nu'} & = & \epsilon'_m \left( \frac{\nu'_c}{\nu'_m}
\right)^{(1-p)/2} \left( \frac{\nu'}{\nu'_c} \right)^{-p/2}, \qquad \nu'_c <
\nu'.
\end{eqnarray}
The definition for fast cooling is analogous (see also
\citealt{Sari1998}). The peak emissivity is given by
\begin{equation}
 \epsilon'_m \backsim \frac{p-1}{2} \frac{\sqrt{3}q_e^3}{m_e c^2}n' B'.
\end{equation}
The synchrotron break frequency $\nu'_m$ depends on its corresponding
critical electron Lorentz factor $\gamma'_m$, leading to
\begin{equation}
 \nu'_m = \frac{3}{4\pi} \frac{q_e}{m_e c} (\gamma'_m)^2 B', \qquad \gamma'_m =
\left( \frac{p-2}{p-1} \right) \frac{\epsilon_e e'_{th}}{n'm_e c^2} .
\end{equation}
For the cooling break frequency an identical relation between
frequency and electron Lorentz factor holds. The critical Lorentz
factor is now given by
\begin{equation}
 \gamma'_c = \frac{ 6 \pi m_e \gamma c}{ \sigma_T (B')^2 t_e},
\end{equation}
which follows from the electron kinetic equation when synchrotron
losses dominate over adiabatic expansion and the cooling time is
approximated by the lab frame time since the explosion.

Synchrotron self-absorption is included in the model using the
assumption that emission and absorption occur in a homogeneous
shell. The solution to the linear equation of radiative transfer then
dictates that we need to replace eq. \ref{flux_model2_equation} by
\begin{equation}
F = \frac{1}{4 \pi d^2_L} \int d \theta d \phi R^2 \sin{\theta} \frac{\epsilon_\nu}{\alpha_\nu} (1 -
\mathrm{e}^{-\tau}),
\end{equation}
The optical depth $\tau \approx - \alpha_\nu \Delta R$ \footnote{This
  is an approximation that does not take into account that not all
  rays cross the homogeneous slab along the radial direction. However,
  significantly increasing the optical depth does not alter our
  finding that self-absorption does not play a role for off-axis VLA
  light curves generated from this model.}. Emissivity and absorption
translate between frames using $\epsilon'_{\nu'} = \gamma^2 (1 - \beta
\mu)^2 \epsilon_\nu$ and $\alpha'_{\nu'} = \alpha_\nu / \gamma (1 -
\beta \mu)$ respectively. In our simplified model we calculate the
absorption coefficient $\alpha'_{\nu'}$ under the assumption that
electron cooling does not influence it. This assumption is justified
when the self-absorption break frequency $\nu_a$ lies well below the
cooling break frequency $\nu_c$, which is the case for all
applications of the model in this paper. Also approximating the
synchrotron spectral shape by just two sharply connected power laws we
then find for the self-absorption coefficient:
\begin{equation}
\alpha'_{\nu'} = (p-1) (p+2) n' \frac{ \sqrt{3} q_e^3 B'}{\gamma'_m 16 \pi m_e^2
c^2} (\nu')^{-2} \left( \frac{ \nu'}{\nu'_m} \right)^{\kappa}, 
\end{equation}
where $\kappa = 1/3$ if $\nu' < \nu'_m$ and $\kappa = -p/2$ otherwise.

Numerically speaking the integration procedure is as follows. First we
tabulate $R(t_e)$ for a given set of physical parameters, so that we
do not need to estimate it analytically but can use its exact
dependence on the fluid Lorentz factor instead. We integrate over
$\theta$ before we integrate over $\phi$, and for each $\theta$,
$\phi$ the angle between observer and fluid element is given by
\begin{equation}
 \mu = \sin{\theta} \cos{\phi} \sin{\theta_{obs}} + \cos{\theta} \cos{\theta_{obs}}.
\end{equation}

Having tabulated $R(t_e)$, we tabulate $\mu( t_e, R(t_e))$ as well for
a given value of $t_{obs}$. Since $\mu(t_e)$ is a monotonically
increasing function of $t_e$, we can unambiguously determine $t_e(
\mu)$ from this table. When determining the value of the integrand at
a given value of $\theta$, $\phi$, we can now calculate the local
fluid conditions and emissivity via $t_e(\mu(\theta, \phi))$.

\bibliography{oa}

\end{document}